\documentclass[final,5p,times,twocolumn]{elsarticle} 

\usepackage{lineno, hyperref}

\usepackage{graphicx}
\usepackage{mhchem}
\graphicspath{images/}
\usepackage{siunitx}
\begin{document}

\begin{frontmatter}

\title{Performance of triple-GEM detectors for the CMS Phase-2 upgrade measured in test beam}

\author[a,b]{Antonello Pellecchia}
\ead{antonello.pellecchia@cern.ch}
\author[b]{Piet Verwilligen} 
\author[a,b]{Anna Stamerra} 
\author{on behalf of the CMS Muon group}
\address[a]{INFN sezione di Bari, Italy}
\address[b]{Università degli studi di Bari, Italy}

\begin{abstract}
    Triple-GEM detectors for the GE2/1 and ME0 stations of the endcap muon system for the Phase-2 upgrade of the CMS Experiment have been operated in a test beam to measure their efficiency and spatial resolution, together with a high spatial resolution triple-GEM tracker. 
     A production module of GE2/1 detectors and a prototype ME0 detector show excellent local efficiency. A prototype detector with GEM foils employing \emph{random hole sectorization} showed significant reduction of dead areas. The spatial resolution of the tracker has been measured and found close to the expected value of \SI{75}{\micro\m}.
\end{abstract}

\end{frontmatter}


\section{Introduction}

In preparation for the High-Luminosity upgrade of the Large Hadron Collider, the Compact Muon Solenoid (CMS) Experiment \cite{cms} is undergoing an upgrade of its muon spectrometer. The Muon upgrade includes the addition of two new stations of triple-GEM detectors, GE2/1 and ME0, to  
increase redundancy for muon track reconstruction in the pseudorapidity region $|\eta|<2.4$ and to extend the muon system coverage to the region $2.4<|\eta|<2.8$.

The Phase-2 CMS GEM detectors have been tested for the first time in their final design with full front-end electronics and final online software. The primary goal is to measure, with high statistics, the efficiency and spatial resolution of the detectors and to demonstrate the operation up to \SI{100}{\kilo\Hz} read-out rate of the front-end ASIC and DAQ. In addition, we measured the performance of a tracking telescope made of triple-GEMs.

The measurements have been made on the H4 line of the CERN North Area. 

\section{The test beam setup}

The test beam setup is made of a tracker of four $10\times 10$ \SI{}{\centi\m\squared} triple-GEM detectors and three detectors under test: a GE2/1 detector, an ME0 detector and a $20\times 10$ \SI{}{\centi\m\squared} triple-GEM prototype. Three scintillators read out by photomultipliers and read out in coincidence by NIM counting modules enclose the setup (one in the front and two in the back) for triggering. The measurements were performed with 80 GeV muons.


\subsection{Triple-GEM detectors}

Each of the four tracking chambers is a $10\times 10$ \SI{}{\centi\m\squared} triple-GEM detector with two perpendicular strip readout planes of \SI{250}{\micro\m} pitch. 
The main detectors under test were a production module of a GE2/1 detector and an ME0 prototype.

The GE2/1 detector has a read-out plane made of radially distributed strips of pitch \SI{952}{\micro\radian} with respect to the nominal interaction point. 
The ME0 detector has strips of pitch \SI{909}{\micro\radian}; 
each GEM foil of the ME0 is divided in sectors along the azimuthal direction with respect to the beam line, a layout choice made to equalize and minimize the gain drop under irradiation in the CMS environment \cite{apellecc_me0}. 
Finally, a special prototype of triple-GEM detector of surface $20\times 10$ \SI{}{\centi\m\squared} was included in the setup. 
The $20\times 10$ prototype has GEM foils divided in two sectors manufactured with a pattern called \emph{random hole sectorization} \cite{florian_random}, 
a technique designed to mitigate the efficiency drops due to the sectorization dead areas.


\subsection{Electronics and DAQ}

All detectors are read out by the VFAT3 front-end ASIC \cite{vfat3}. In the GE2/1 and ME0 detectors the signals induced on the readout strips are routed to the VFATs through a GEM Electronics Board (GEB) attached to the readout plane; the GEB also routes the power lines to the VFATs and contains the electronic connections between the VFATs and the OptoHybrid (OH) board. The OH provides slow control to the VFATs through GBT \cite{gbt} ASICs and sends the VFAT data to the back-end through the VTRx \cite{vtrx} chip. The tracking detectors are read out by two additional GE2/1 GEBs.

The back-end is made of a commercial PCIe board (CVP-13) based on a Xilinx UltraScale+ VU13P FPGA. The communication between the back-end and the OptoHybrid VTRx is estabilished through optical links with QSFP transceivers. The trigger signal from the scintillators is received by the FPGA through one pin of the CVP-13 DIMM card and is sent to the front-end after being synchronized to the \SI{40}{\mega\Hz} clock generated by the back-end board. The raw DAQ data received by the back-end are sent via an additional optical link to a network interface card on the DAQ machine and stored locally for reconstruction.

\section{Reconstruction and analysis workflow}

Before the analysis, the raw data undergo a reconstruction up to track building. After the strip information of the hip event-by-event has been extracted from the binary data, clusters of neighbouring strips are grouped together to form reconstructed hits or \emph{rechits}; the uncertainty of each reconstructed hit is set to $\text{cluster size}/\sqrt{12}$. A track trajectory is fitted from the the tracker rechits and the fit is used to interpolate the track positions to the GE2/1 and ME0 detector positions. 

At the beginning of the analysis, an iterative software alignment based on rotation and translation corrections has been performed on all detectors.


\section{Results}

\begin{figure}[h]
    \centering
    \includegraphics[width=\linewidth]{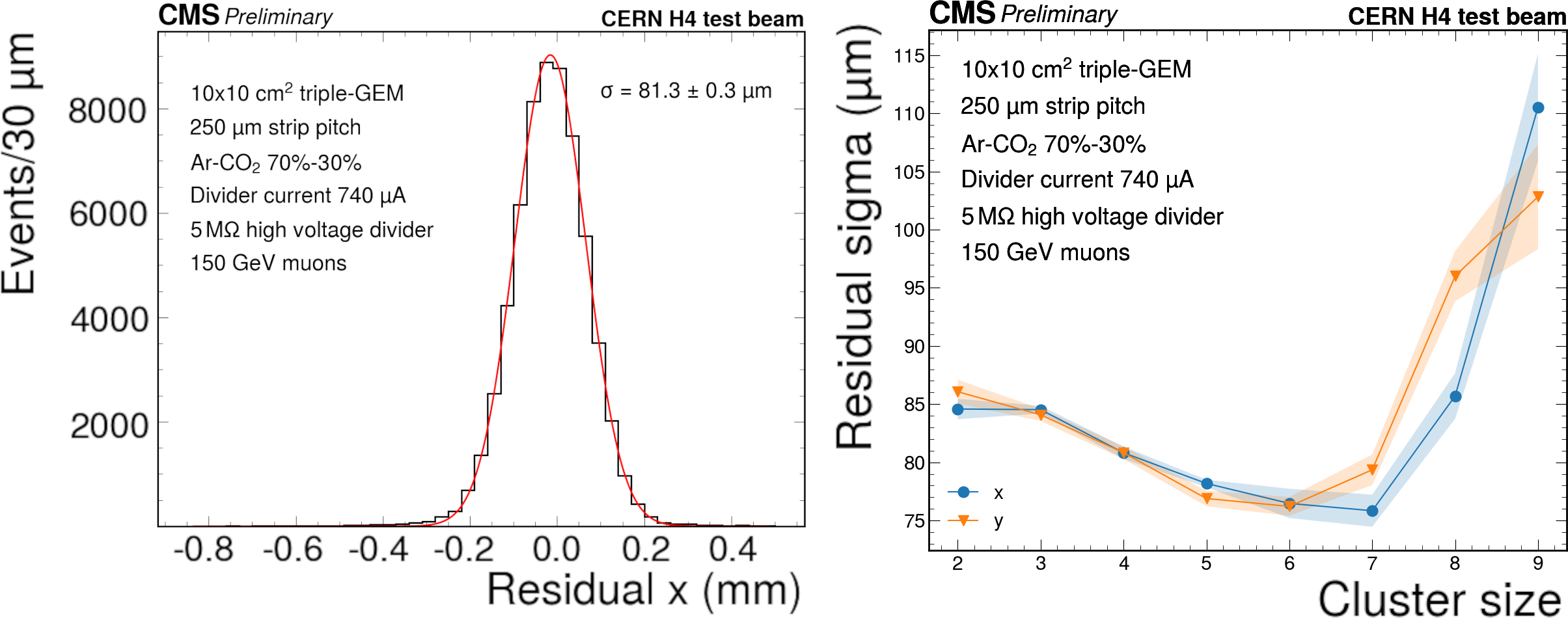}
    \caption{(Left) residuals of tracker 2 for all cluster sizes. (Right) Dependence of spatial resolution on the cluster size for the same detector.}
    \label{fig:tracker_plots}
\end{figure}

The spatial resolution and efficiency of each tracking detector were measured after removing it from the track building algorithm. Figure \ref{fig:tracker_plots} left shows that the average spatial resolution of tracker 2 is \SI{81.3+-0.3}{\micro\m}, close to the expected value of \SI{75}{\micro\m} (equal to strip pitch$/\sqrt{12}$); as shown in Fig.~\ref{fig:tracker_plots} right, the spatial resolution is strongly dependent on the event cluster size, with best resolution between 4 and 7 strips. 
Worse resolution at high cluster sizes may be due to delta rays from the avalanche; 
worse resolution at low cluster size can be associated to strips falling below the VFAT threshold and being excluded from the cluster, thus biasing the cluster center position. 
The average efficiency of each tracking detector was 92\% and likely limited by the high front-end thresholds set due to the high electronic noise (on average between 0.5 and \SI{1}{\femto\coulomb} for all VFATs).

\begin{figure}[h]
    \centering
    \includegraphics[width=\linewidth]{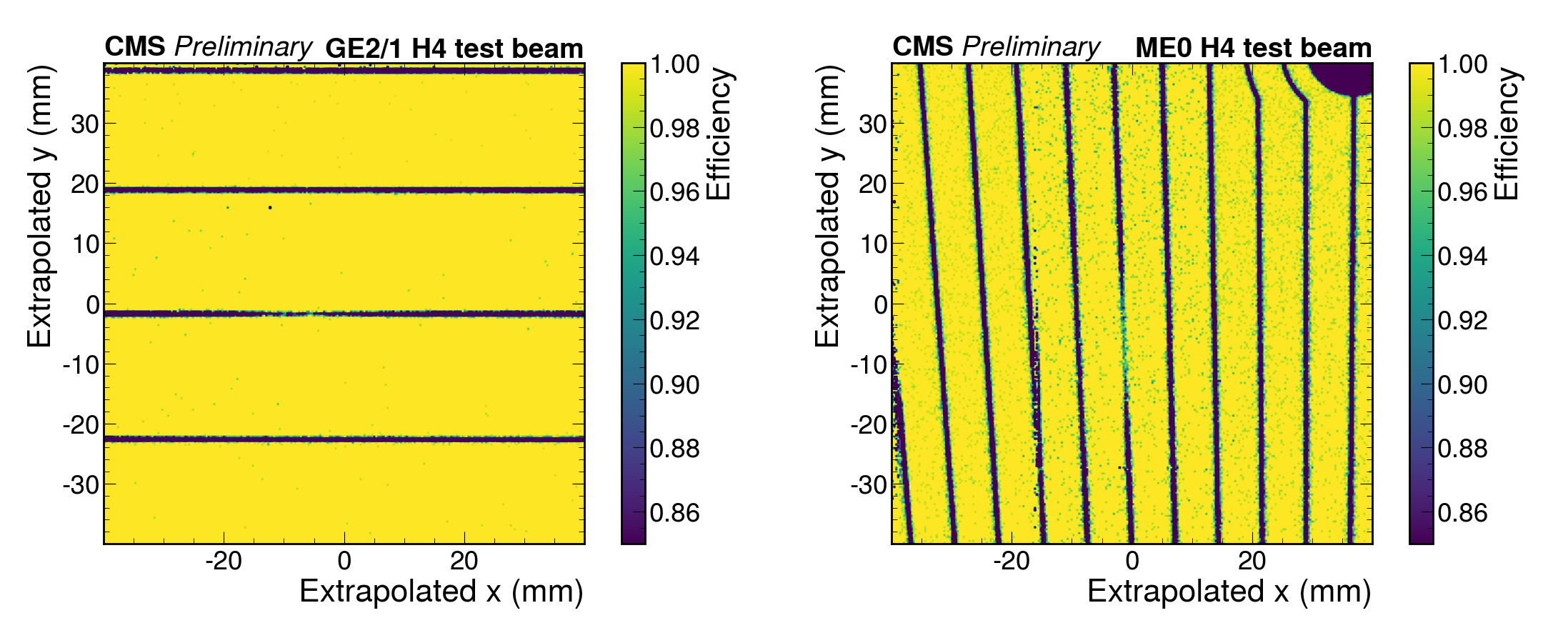}
    \caption{Efficiency maps of the GE2/1 and ME0 detectors, showing the efficiency drops corresponding to the GEM foil sectorization.}
    \label{fig:efficiencies}
\end{figure}

Efficiency maps of the GE2/1 and ME0 detectors (Fig.~\ref{fig:efficiencies}) show that both detectors have large regions with efficiency close to 100\%. Average detector efficiencies are lower (98\% for the GE2/1 detector and 96.2\% for ME0) because of the dead areas due to the GEM foil sectorization. The excellent efficiency of the Phase-2 detectors could be obtained because of the low VFAT thresholds, due to the low electronic noise (below 0.3 fC) with respect to the tracker achievable thanks to the GEB shielding and the redundant VFAT3 plugin card grounding connections.

\begin{figure}[h]
    \centering
    \includegraphics[width=0.62\linewidth]{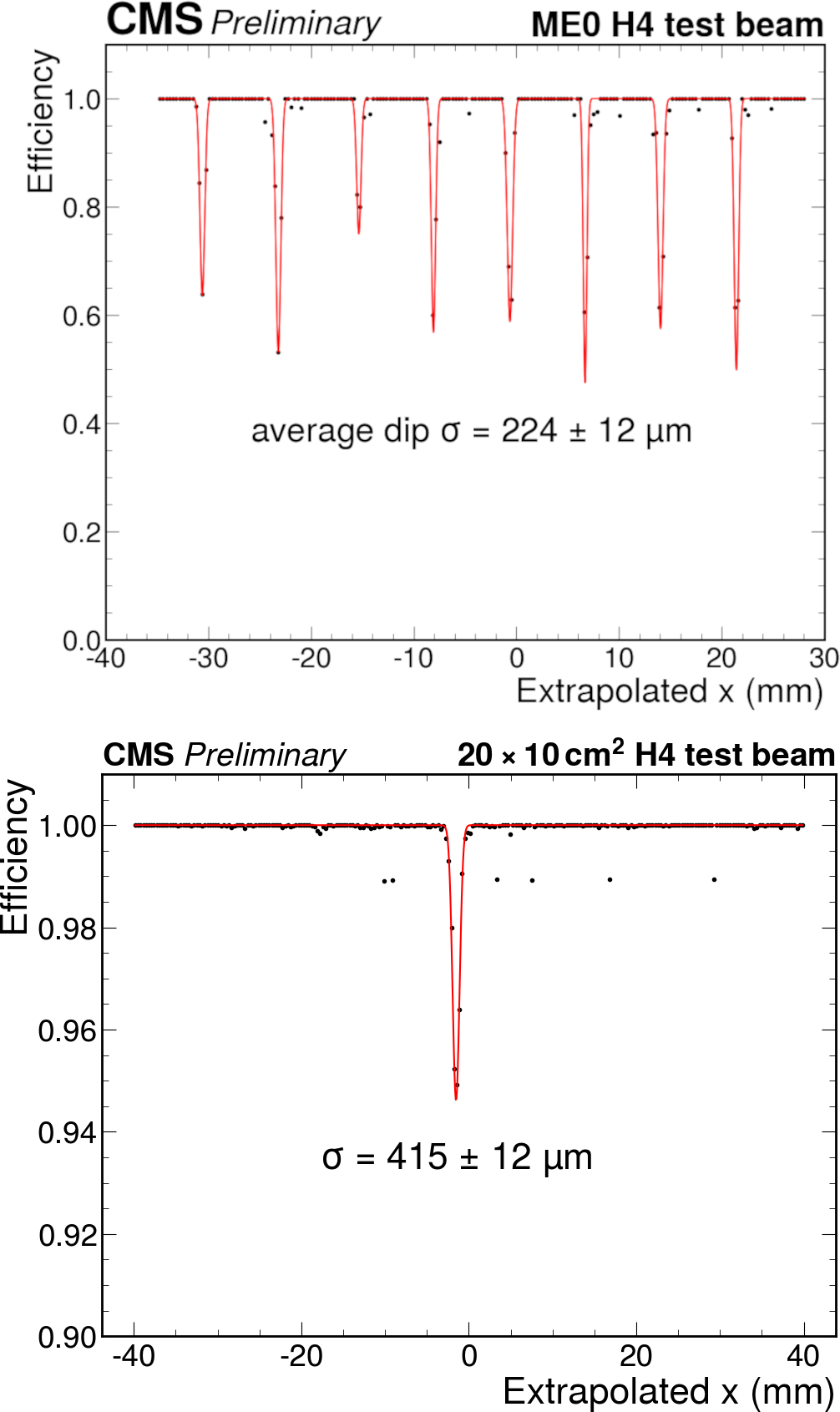}
    \caption{Efficiency profiles along x of the ME0 and random-segmented detectors for a slice in y. The profiles are fit with a gaussian comb and the width of the efficiency dip is the $\sigma$ of the gaussian fit.}
    \label{fig:efficiencies_slice}
\end{figure}

Figure~\ref{fig:efficiencies_slice} shows the efficiency profile in the x direction for a slice along y for two detectors: the ME0 and the $20\times 10$ random-hole sectorized prototype. While the ME0 detector shows regions with efficiency drops up to 50\%, the $20\times 10$ prototype has an efficiency reduction of only 4\% in the sector border region. However, the width of the efficiency dip in the $20\times 10$ is almost double the average dip width on the ME0.

\section{Conclusion}

The test beam has successfully proven the good performance of the CMS Phase-2 triple-GEM detectors, with excellent muon detection efficiency. The random hole sectorization technique has shown the possibility to reduce the dead area due to the GEM foil sectorization and is under consideration for future ME0 R\&D.


\smallskip

\end{document}